\documentclass[conference]{IEEEtran}
\IEEEoverridecommandlockouts
\usepackage[nospace,noadjust]{cite}
\usepackage[english]{babel}
\usepackage[dvipsnames]{xcolor}
\usepackage{amsmath,amssymb,amsfonts}
\usepackage{amsthm}
\usepackage{graphicx}
\usepackage{bbm}
\usepackage{setspace}
\usepackage{braket}
\usepackage{algpseudocode}
\usepackage[]{algorithm}
\usepackage{textcomp}
\usepackage{xcolor}
\def\BibTeX{{\rm B\kern-.05em{\sc i\kern-.025em b}\kern-.08em
		T\kern-.1667em\lower.7ex\hbox{E}\kern-.125emX}}
\usepackage{nccmath}
\usepackage{array}
\usepackage{dutchcal}
\usepackage{booktabs}
\definecolor{wblue}{RGB}{0,58,199}
\usepackage{setspace}
\usepackage[T1]{fontenc}
\usepackage[utf8]{inputenc}
\usepackage[font=small,labelfont=bf]{caption}
\usepackage{bm}
\algnewcommand{\IIf}[1]{\State\algorithmicif\ #1\ \algorithmicthen}
\algnewcommand{\EndIIf}{\unskip\ \algorithmicend\ \algorithmicif}
\algnewcommand{\FFor}[1]{\State\algorithmicfor\ #1\ \algorithmicdo}
\algnewcommand{\EndFFor}{\unskip\ \algorithmicend\ \algorithmicfor}

\ifCLASSINFOpdf
\else
\fi
\begin{document}
	%
	% paper title
	% Titles are generally capitalized except for words such as a, an, and, as,
	% at, but, by, for, in, nor, of, on, or, the, to and up, which are usually
	% not capitalized unless they are the first or last word of the title.
	% Linebreaks \\ can be used within to get better formatting as desired.
	% Do not put math or special symbols in the title.
	\title{
		{{Sum Capacity Maximization in Multi-Hop Mobile Networks with Flying Base Stations}}
		{\footnotesize \textsuperscript{}}}
	
	\author{\IEEEauthorblockN{Mohammadsaleh Nikooroo\textsuperscript{1}, Omid Esrafilian\textsuperscript{2}, Zdenek Becvar\textsuperscript{1}, David Gesbert\textsuperscript{2}}
		\IEEEauthorblockA{\textit{\textsuperscript{1} Faculty of Electrical Engineering} 
			\textit{Czech Technical University in Prague},
			Prague, Czech Republic \\
			\textsuperscript{2}	\textit{Communication Systems Department, EURECOM}, Sophia Antipolis, France\\
			\textsuperscript{1}\{nikoomoh,zdenek.becvar\}@fel.cvut.cz, \textsuperscript{2}\{esrafili,gesbert\}@eurecom.fr }
		%}
		
		%\author{Mohammadsaleh Nikooroo,~\IEEEmembership{Member,~IEEE,}
		%	Zdenek Becvar,~\IEEEmembership{Senior Member, ~IEEE}

		% }{Sum Rate Maximization in UAV-Assisted Mobile Networks}
		%
		% author names and IEEE memberships
		% note positions of commas and nonbreaking spaces ( ~ ) LaTeX will not break
		% a structure at a ~ so this keeps an author's name from being broken across
		% two lines.
		% use \thanks{} to gain access to the first footnote area
		% a separate \thanks must be used for each paragraph as LaTeX2e's \thanks
		% was not built to handle multiple paragraphs
		%
		
		%\author{Mohammadsaleh Nikooroo,~\IEEEmembership{Member,~IEEE,}
		%	Zdenek Becvar,~\IEEEmembership{Senior Member, ~IEEE}
		% <-this % stops a space
		%\thanks{Mohammadsaleh Nikooroo and Zdenek Becvar are with Department of Telecommunication Engineering, Faculty of Electrical Engineering, Czech Technical University in Prague, Czech Republic, e-mail: (nikoomoh@fel.cvut.cz, and zdenek.becvar@fel.cvut.cz).}
		\thanks{\textcolor{black}{This work was supported by the project No. LTT 20004 funded by Ministry of Education, Youth and Sports, Czech Republic and by the grant of Czech Technical University in Prague No. SGS20/169/OHK3/3T/13, and partially by the HUAWEI France supported Chair on Future Wireless Networks at EURECOM.}}}% <-this % stops a space
	%\thanks{J. Doe and J. Doe are with Anonymous University.}% <-this % stops a space
	%\thanks{Manuscript received April 19, 2005; revised August 26, 2015.}}

	% make the title area
	\maketitle
	
	% As a general rule, do not put math, special symbols or citations
	% in the abstract or keywords.
	\begin{abstract}
		Deployment of multi-hop network of unmanned aerial vehicles (UAVs) acting as flying base stations (FlyBSs) presents a remarkable potential to effectively enhance the performance of wireless networks. Such potential enhancement, however, relies on an efficient positioning of the FlyBSs as well as a management of resources. In this paper, we study the problem of sum capacity maximization in an extended model for mobile networks where multiple FlyBSs are deployed between the ground base station and the users. Due to an inclusion of multiple hops, the existing solutions for two-hop networks cannot be applied due to the incurred backhaul constraints for each hop.  To this end, we propose an analytical approach based on an alternating optimization of the FlyBSs’ 3D positions as well as the association of the users to the FlyBSs over time. The proposed optimization is provided under practical constraints on the FlyBS’s flying speed and altitude as well as the constraints on the achievable capacity at the backhaul link. The proposed solution is of a low complexity and extends the sum capacity by 23\%-38\% comparing to state-of-the-art solutions. 
	\end{abstract}

	
	\section{Introduction}\label{sec:1}
	\par
Unmanned aerial vehicles (UAVs) have attracted an abundance of research interest in wireless communications in the last few years thanks to their high mobility and adaptability to the environment. Deployed as flying base stations (FlyBSs), UAVs can potentially bring a great improvement in applications such as surveillance, emergency situations, or providing user’s coverage in areas with unreliable connectivity \cite{Liu2021}, \cite{Nikooroo2022TNSE}, \cite{Li2019},\cite{Li2020}. Several challenges exist to enable an effective use of FlyBSs, including an efficient cooperation between the FlyBSs’ via a management of the resources as well as FlyBSs’ positioning. An important case with cooperative FlyBSs is relaying networks where FlyBSs either  serve the ground users directly (access link) or relay the data to establish a connection between the users and the ground base station (GBS).
 
	Several recent works target enhancing the performance in networks with relaying FlyBSs. With respect to those works only focusing on the communication at the access link, relaying networks necessitate to consider the backhaul link connecting the users to the GBS. In particular, flow conservation constraints apply at each relay node to ensure a sufficient backhaul capacity for the fronthaul link. 
	The basic model for relaying FlyBS networks is a two-hop architecture where all FlyBSs directly serve users at the access link and also connect directly to the GBS via the backhaul link. A majority of recent works target an enhancement in two-hop relaying networks with a consideration of backhaul. 
	
	The problem of resource allocation and FlyBS's positioning is considered in many works targeting various objectives, including optimization of minimum rate for delay-tolerant users \cite{Huang2020},  energy consumption \cite{Qiu2020Access},  network profit gained from users \cite{Cicek2020}, sum capacity \cite{Pham2021}, network latency \cite{Yu2021}. 
	The mentioned works \cite{Huang2020}-\cite{Yu2021} consider a single FlyBS, and an application of those works to multiple-FlyBS scenario is not trivial. 
	
	Several works also consider multiple FlyBSs in two-hop relaying networks. In \cite{Qiu2020TCOM} the authors study a joint placement, resource allocation, and user association of FlyBSs to maximize the network’s utility. %
	Furthermore, the authors in \cite{Mach2022} maximize the sum capacity via FlyBS’s positioning, user association, and transmission power allocation. In \cite{Iradukunda2021} the minimum rate of the users is maximized via resource allocation and positioning in wireless backhaul networks. Furthermore, the authors in \cite{Pan2019} investigate an optimization the FlyBS’s position, user association, and resource allocation, to maximize the utility in software-defined cellular networks with wireless backhaul. 
	Due to the introduced flow conservation constraints, an extension of studies/solutions on two-hop FlyBS networks to higher number of hops is often not simple or straightforward. There are quite a limited number of works that consider relaying FlyBSs in networks with more than two hops. In \cite{Li2018} the minimum downlink throughput is maximized by optimizing the FlyBSs’ positioning, bandwidth, and power allocation. The provided solution, however, does not address interference management as orthogonal transmissions is assumed. Furthermore, the FlyBSs’ altitude is not optimized.
	 Then, in \cite{Sabzehali2021} the number of FlyBSs is optimized while ensuring both coverage to all ground users as well as backhaul connectivity to a terrestrial base station. The authors in \cite{Wang2020} investigate an interference management scheme based on machine learning and a positioning based on K-means to mitigate interference and FlyBSs’ power consumption.
	
	In the view of existing works on relaying FlyBS networks, we are motivated to take one step forward and to address a maximization of sum capacity via a placement of FlyBSs and an association of users in a multi-hop relaying FlyBS architecture where the FlyBSs serving the users at the access link connect to a GBS via another relaying FlyBS. Such an extension from two-hop model would allow a vaster range of user coverage to connect more remote users to the GBS. Unlike the most of related works, in our model, also the GBS and the relay are allowed to serve the users directly. In contrast to most of related works, a reuse of channels from the access link is enabled to establish the backhaul connection. The solution is provided under backhaul constraints.  
	
	The main contribution of this paper is explained as follow. We provide a framework based on a multi-hop FlyBS wireless network where the FlyBSs at the access link communicate with a ground base station through a relaying FlyBS. We formulate the network’s sum capacity with a consideration of channel resue for the backhaul link. We formulate the problem of sum capacity maximization via an association of the users and a positioning of the FlyBSs at the access link and the relay. In our model, a direct serving of the users by the relaying FlyBS as well as by the GBS is also possible. A heuristic iterative solution is proposed based on an alternating optimization of the FlyBSs’ positions at the access link, FlyBS’s position at the relay, and then a reassociation of the users to the FlyBSs. An approximation of the sum capacity is proposed to derive a radial function to determine the FlyBSs’ optimal directions of movement in the proposed iterative positioning. 
	
	The rest of this paper is organized as follows. In Section II we elaborate the system model for multi-hop FlyBS network.  Next, the problem of sum capacity maximization is formulated and our proposed solution to the FlyBS’s positioning and user association is provided in  Section III. Then, in section IV, we specify our adopted simulation scenario and we show the performance of our proposed solution and we compare it with existing works. Last, we conclude the paper and outline the potential extensions for the future work.

	%\vspace{-1\baselineskip}
	
	\section{System model and problem formulation}\label{sec:2}
	
In this section, we define the system model and provide details about  transmission power and channel capacity.

	We consider a set of $ M $ FlyBSs and a ground base station (GBS) serving $ N $ ground users. $ M-1 $ of those FlyBSs serve at the access link. The backhaul communication between those $ M-1 $ FlyBSs and the GBS is established via an intermediate relay FlyBS. Fig. 1 illustrates the adopted model. 
	Let $ \bm{Q}=\{\bm{q_1},…,\bm{q_M}\} $ be the set of the FlyBS’s positions where $ \bm{q_m} [k]= $$ [X_m [k],Y_m [k],H_m [k]]^T $ denote the location of the $m$-th FlyBS at the time step $ k $ $ (1\leq m\leq M) $, where the index $ m=M $ indicates the relay. Let $ \bm{q_{M+1}}= $ $ [X_{M+1},Y_{M+1},H_{M+1}]^T $ denote the GBS’s position. Next, let $ d_{m_1,m_2 } [k] $ denote the Euclidean distance between the $ m_1 $-th and $ m_2 $-th BSs’ receivers (we use the general term BS when referring to both GBS and FlyBSs). Furthermore, let $ v_n [k]= $$  [x_n [k],y_n [k],z_n [k]]^T $ denote the coordinates of the $n$-th ground user at the time step $k $. Then, $ d_{n,m}  [k] $ denotes Euclidean distance of the $n$-th user to the $m$-th BS. 
	As in many related works, we assume that the current positions of the users are known to the BSs. Also, the FlyBSs can determine their own position \cite{Pham2021}, \cite{Qiu2020TCOM}, \cite{Li2018}, \cite{Esrafilian}. 
	Let $ \textbf{A}=(a_{n,m}) \in {\{0,1\}}^{N\times(M+1)}  $be the association matrix where $ a_{n,m}= $1  indicates an association of the $ n $-th user to the $ m $-th BS. Note that the users can be directly served by the relay or the GBS as well. Every user cannot be associated to more than one BS. 
	Also, we assume the whole radio band is divided into the set of channels $ \textbf{L}=\{l_1,…,l_C\} $, where channel $ l_c $ has a bandwidth of $ B_c $ $ (1\leq c\leq C) $. Note that the channels can be of different bandwidth in our model.  We adopt orthogonal downlink channel allocation for all users associated to the same BS. Furthermore, let $ g_n $ be the index of the channel allocated to the $ n $-th user. Also, we assume $ I_M  $ and $ I_{M+1} $ denote the set of indices of channels allocated to the users served by the relay and by the GBS, respectively. Also, let $ I_{M,m} $ be the set of channels’ indices used between the relay and the $ m $-th FlyBS at the access link. The relay communicates with users and other FlyBSs using orthogonal channels. Note that, we do not target an optimization of channel allocation due to space limit, and we leave that for future work. Nevertheless, our model works with any channel allocation.

The received power from the $m$-th FlyBS at the $n$-th user is denoted as $ p_{n,m}^R $ and calculated as:
\small
	\begin{gather}
	p_{n,m}^R=\Gamma_{n,m} (\frac{\gamma}{\gamma+1}\overline{h}_n +\frac{1}{\gamma+1}\tilde{h}_n ) d_{n,m}^{-\alpha_{n,m} }=
	Q_{n,m} d_{n,m}^{-\alpha_{n,m} },    
	\end{gather}%\vspace{-1\baselineskip}
	\normalsize	
	
	\noindent where $ \Gamma_{n,m} $ is a parameter depending on communication frequency and gain of antennas. Furthermore, $ \gamma $ is the Rician fading factor, $ \overline{h}_n $ is the line-of-sight (LoS) component satisfying $ |h_n |= $1, and $\tilde{h}_n$ denotes the non-line-of-sight (NLoS) component satisfying $ \tilde{h}_n \sim $ $ CN(0,1) $, and $\alpha_{n,m}$ is the pathloss exponent. Note that the coefficient $ \Gamma_{n,m} (\frac{\gamma}{\gamma+1}\overline{h}_n +\frac{1}{\gamma+1}\tilde{h}_n ) d_{n,m}^{-\alpha_{n,m} } $ is substituted with $ Q_{n,m} $ for an ease of presentation in later discussions. Similar relation applies for backhaul link as $ p_{m_1,m_2,k}^R= $$ Q_{m_1,m_2,k} d_{m_1,m_2} ^{-\alpha_{m_1,m_2}} $ where $ p_{m_1,m_2,k}^R $ is the received power at $ m_1 $-th BS from  $ m_2 $-th BS over $ k $-th channel.  
	\begin{figure}[!t]
		\centering
		\includegraphics[width=3.42in]{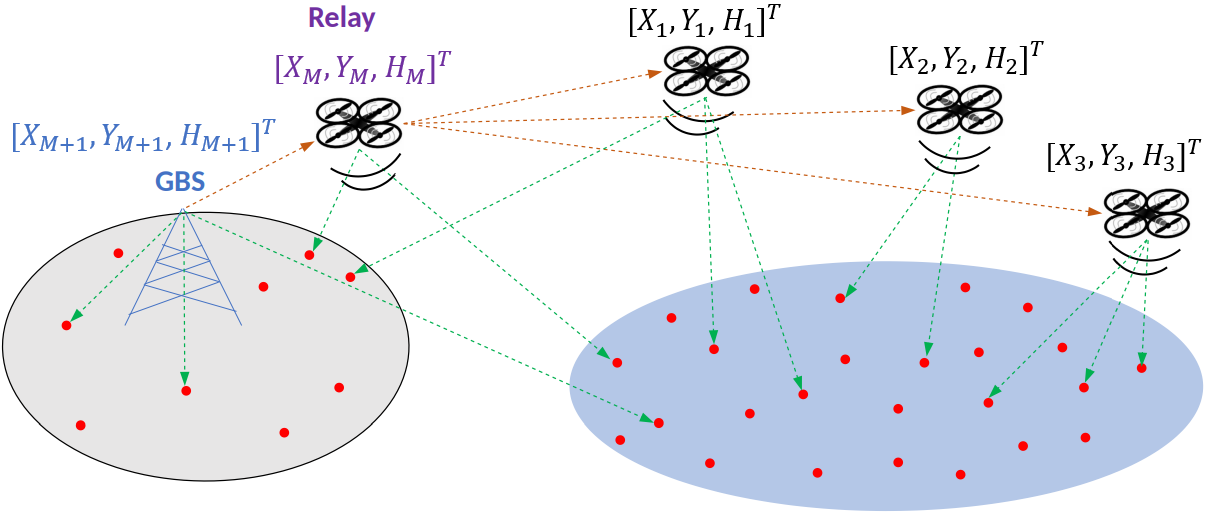}	
		\captionsetup{justification=centering}
		\caption{System model with the FlyBSs at the access link, relaying FlyBS, and the GBS serving moving users.} 
		\label{fig:sysmodel}
	\end{figure}\vspace{0\baselineskip}
	
	The downlink capacity of the $n$-th user is calculated as	
	\small
	\begin{gather}
		{C_{n,m}=a_{n,m} B_{g_n}  \text{log}_2(1+\frac{p_{n,m}^R}{\sigma_{n,m}^2+\sum_{m'\in \{a_{n,m'}=0\}} p_{n,m'}^R})}
	\end{gather}%\vspace{-1\baselineskip}
\normalsize
\noindent where $ \sigma_{n,m}^2 $ is the noise power. Next, the capacity between the relay and the $m$-th FlyBS is 
\small
	\begin{gather}
	C_{M,m}=\sum_{k\in I_{m,M}} B_k  \text{log}_2(1+
	\frac{p_{m,M,k}^R}{\sigma_{m,M,k}^2+\sum\limits_{\substack{m'\in [1,M+1]\backslash{\{M\}}}} p_{m,m',k}^R})
\end{gather}%\vspace{-1\baselineskip}
\normalsize

\noindent Also, the link's capacity between the GBS and the relay is  
\small
	\begin{gather}
	C_{M+1,M}=\sum_{k\in [1,K]\backslash I_{M+1}} B_k  \text{\text{log}}_2(1+
	\frac{p_{M,M+1,k}^R}{\sigma_{M,M+1,k}^2+\sum_{m=1}^{M-1} p_{M,m,k}^R})
\end{gather}%\vspace{-1\baselineskip}
\normalsize

In the next section we formulate the problem of sum capacity maximization and we elaborate our proposed solution.

	%\vspace{-1\baselineskip}
	
	\section{Problem Formulation and Proposed FlyBS positioning and User Association}\label{sec:3}
	
	%\vspace{-1\baselineskip}

In this section, we first introduce the problem of sum capacity maximization. Then, we outline our proposed optimization of user association and FlyBS’s positioning.

\subsection{Problem Formulation}\label{ssec:3A}

The objective is to find a 3D positioning of the FlyBSs as well as an association of users to the BSs (including GBS and relay) to maximize the sum capacity at every time step $ k $ while constraints on the FlyBSs’ altitude and speed as well as on backhaul are taken into account. Hence, we formulate the problem of the sum capacity maximization as follows:
\begin{gather}
		\label{eqn:problem_formulation}
		\operatorname*{max}_{{\bm{Q}},\textbf{A}} \sum_{m=1}^{M}\sum_{n=1}^{N} a_{n,m}C_{n,m}[k],\quad \forall k\\\nonumber
		s.t.\quad H_{min,m} [k]\leq H_m[k]\leq H_{max,m}[k],\quad (5a)\nonumber\\
		V_{F,m} [k]\leq V_m^{max}, \quad \forall m\in [1,M] \quad\quad\quad (5b) \nonumber\\[-6pt]
		\sum_{n=1}^{N} a_{n,m}C_{n,m}[k]\leq C_{M,m}[k],  m\in [1,M-1] \quad (5c)\nonumber\\[-9pt]
		\sum_{m=1}^{M-1} C_{M,m}[k]\leq C_{M+1,M}[k], \quad\quad\quad (5d)\nonumber\\[-8pt]
		a_{n,m} [k] \in\{0,1\},\sum_{m=1}^{M+1} a_{n,m}[k]\leq 1,  \quad (5e)\nonumber\\[-11pt]
		%	\sum_{m=1}^{M+1} a_{n,m}[k]\leq 1,    \quad\quad\quad\quad\quad (5f)\nonumber\\[-8pt]
			\sum_{n=1}^{N} a_{n,m}[k]\leq C, \quad\forall m\in [1,M-1]   \quad  (5f)\nonumber\\[-10pt]
			\sum_{n=1}^{N} a_{n,m}[k]\leq |I_m|, \quad\forall m\in \{M,M+1\},   \quad\quad (5g)\nonumber
\end{gather}
where $ H_{min,m} $ and $ H_{max,m} $ denote the minimum and maximum flying altitude of the $m$-th FlyBS, respectively, and are determined with respect to the topology of the environment and the flying regulations. Furthermore,  $ V_m^{max} $ is the $m$-th FlyBS’s maximum supported speed. The constraints (5a) and (5b) always ensures a flight within the allowed range of altitude and speed, respectively. The constraint (5c) guarantees that the backhaul link’s capacity between each FlyBS at the access link and the relay is larger than the sum downlink capacity of that FlyBS, and the constraint (5d) implies that the capacity of the GBS-to-relay link is larger than the sum capacity for the links between relay and FlyBSs at the access link.  The constraint (5e) indicates that each user is not associated to more than one BS, and the constraints (5f) and (5g) ensure that the number of users associated to each BS cannot exceed the number of channels allocated to each BS. 

Challenges regarding the optimization problem in (5) include: 1) the sum capacity function is non-convex with respect to the FlyBSs’ positions (i.e., \bm{$ {q_m} $}, $ m\in [1,M] $), 2) the constraints (5c) and (5d) are non-convex with respect to the FlyBSs’ positions, and 3) the discrete association function $ \textbf{A} $ in (5) makes the optimization problem non-tractable. 
To tackle the challenges mentioned above we propose a heuristic solution by the means of approximation and based on converting the objective to a radial function to determine FlyBSs' movement towards an increase in the sum capacity. Then, the proposed solution to (5) is provided based on an alternating optimization of the FlyBSs’ positions at the access link, relay’s position, and user association. In particular, for a given user association, we propose an iterative approach based on an optimization of positioning of the FlyBSs at the access link under the constraint on their backhaul link and other constraints on their movement. Then, a positioning of the relay is proposed under the constraint on backhaul links between the relay and other BSs.  Then, a reassociation of the users to BSs at their updated positions is applied.  

\subsection{Approximation of sum capacity as radial function}\label{ssec:3B}

	To proceed with the solution, we first propose and derive an approximation of the objective (sum capacity)that converts the objective to a radial function indicating the direction of movement for all FlyBSs to maximize the sum capacity. To begin with, we rewrite the logarithm term in (2) using (1) as

	\small
	\begin{gather}
		\label{apx:1}
		%\text{log}_2(1+\frac{Q_{n,m}d_{n,m}^{-\alpha_{n,m}}}{\sigma_{n,m}^2+\sum_{m'\in \{a_{n,m'}=0\}} Q_{n,m'}d_{n,m'}^{-\alpha_{n,m'}}})=\nonumber\\
		\text{log}_2(1+\frac{\frac{Q_{n,m}}{\sigma_{n,m}^2}d_{n,m}^{-\alpha_{n,m}}}{1+\sum_{m'\in \{a_{n,m'}=0\}} \frac{Q_{n,m'}}{\sigma_{n,m}^2}d_{n,m'}^{-\alpha_{n,m'}}})=
		\text{log}_2(1+\nonumber\\\frac{Q_{n,m}}{\sigma_{n,m}^2}(d_{n,m}^2)^{-\frac{\alpha_{n,m}}{2}}+\sum_{m'\in \{a_{n,m'}=0\}} \frac{Q_{n,m'}}{\sigma_{n,m}^2}(d_{n,m'}^2)^{-\frac{\alpha_{n,m'}}{2}})\nonumber\\-
		\text{log}_2(1+\sum_{m'\in \{a_{n,m'}=0\}} \frac{Q_{n,m'}}{\sigma_{n,m}^2}(d_{n,m'}^2)^{-\frac{\alpha_{n,m'}}{2}}).
	\end{gather}
	\normalsize
	
	\noindent Then, we use the first-order Taylor approximation $ \text{log}_2(1+X)\approx \frac{X}{\text{log}(2)} $ to expand the logarithm terms on the right-hand side in (6) and we get a linear expression with respect to $(d_{n,m}^2)^{-\frac{\alpha_{n,m}}{2}} $. Next, we rewrite the term $(d_{n,m}^2)^{-\frac{\alpha_{n,m}}{2}} $ as 
	
	\small
	\begin{gather}
		(\Psi^2+((X_m-x_n)^2+(Y_m-y_n)^2+(H_m-z_n)^2)-\Psi^2)^{-\frac{\alpha_{n,m}}{2}},
	\end{gather}
	\normalsize
	where $ \Psi $ is an arbitrary nonzero constant used to expand (7) as follows. The right-hand side in (7) is in the form of $ (a+\chi)^k $ where $ k={\frac{\alpha_{n,m}}{2}} $, $ a=\Psi^2 $, and $ \chi=((X_m-x_n)^2+(Y_m-y_n)^2+(H_m-z_n)^2-\Psi^2) $. Using the first-order Taylor approximation with respect to $ \chi $,  (7) is  converted to a summation of linear terms with respect to $\chi$. Next, suing the equation  $ 	\sum_{n=1}^{N}\beta_n(X_m-x_n)^2 =$ $(\sum_{n=1}^{N}\beta_n)(X_m-\frac{\sum_{n=1}^{N}\beta_nx_n}{\sum_{n=1}^{N}\beta_n})^2+(\sum_{n=1}^{N}\beta_nx_n^2-
	\frac{(\sum_{n=1}^{N}\beta_nx_n)^2}{\sum_{n=1}^{N}\beta_n})$ for any weighted sum of squares (here shown only for $ X_m $), the sum capacity can be approximated as
	%\small
		\begin{gather}
		\sum_{n=1}^{N}C_{n,m}[k]\approx\zeta[k] -\sum_{1\leq m\leq M}\rho_m||\bm{q_m}[k]-\bm{q_{0,m}}[k]||^2.
	\end{gather}
\normalsize

\noindent	where $ \zeta $, $ \rho_m $, and $ \bm{q_{0,m}} $ are constants with respect to   $ \bm{q_m} $.
	
	Having the sum capacity as presented in (8), it is seen that for $ \rho_m>0 $ ($ \rho_m<0 $) a movement of the $m$-th FlyBS towards (against)  $ \bm{q_{0,m}} $ causes an increase in the sum capacity. This fact helps to determine the FlyBS’s positions under the constraints (5a)-(5d) as elaborated in the next subsection.

	\subsection{Positioning of the FlyBSs at the access link }\label{ssec:3C}

	The approximation in (8) is exploited to reposition the FlyBSs at the access link (repositioning of relay will be addressed separately in the next subsection). First, we relax the GBS-to-relay constraint (5d) and we propose a positioning of the FlyBSs according to the constraints (5a)-(5c). 
	The constraint (5a) limits the position of the $m$-th FlyBS ($ q_m $) between the planes $ z=H_{min,m} [k] $ and $ z=H_{max,m} [k] $. Then, the constraint (5b) bounds $ q_m $ to the points on or inside of a sphere with a center at $ q_m  [k-1]$ and with a radius of $  V_m^{max}\delta $ where $\delta$ is the time distance between two consecutive time steps. 
	Regarding the constraint (5c), there exist terms in $ C_{n,m} $ and $ C_{M,m} $ (in (2) and (3)) related to the interference from other FlyBSs that complicates a dealing with (5c) as (5c) defines a non-convex region with respect to $ q_m $. To tackle this issue, we use the fact that the FlyBSs’ movements are limited at each time step due to the limited speed. We convert (5c) into a convex constraint in following way. First, let $ d_{n,m,min} [k] $  and $ d_{n,m,max} [k] $ denote the minimum and maximum distances that can occur between the $n$-th user and the $m$-th FlyBS at the time step $ k $, respectively. $ d_{n,m,min} [k] $    and $ d_{n,m,max} [k] $  are calculated using the FlyBSs' speeds as:

	\begin{gather}
	d_{n,m,min}=||\bm{q_m} [k-1]-v_n [k]||-V_m^{max} \delta,\nonumber\\
	d_{n,m,max}=||\bm{q_m} [k-1]-v_n [k]||+V_m^{max} \delta.
	\end{gather}

	\noindent Hence, the left-hand side in (5c) is upper bounded by
	
	%\sum_{n=1}^{N}a_{n,m} B_{g_n}  \text{log}_2(1+\frac{p_{n,m}^R}{\sigma_{n,m}^2+\sum_{m'\in \{a_{n,m'}=0\}} p_{n,m'}^R})\leq \nonumber\\
	\small
	\begin{gather}
			\sum_{n=1}^{N}a_{n,m} B_{g_n}  \text{log}_2(1+\frac{Q_{n,m}d_{n,m,min}^{-\alpha_{n,m}}}{\sigma_{n,m}^2+\sum_{m'\in \{a_{n,m'}=0\}} Q_{n,m'}d_{n,m',max}^{-\alpha_{n,m'}}})\label{LHS-6c}
	\end{gather}%\vspace{-1\baselineskip}
		\normalsize	
		
	\noindent Next, the lower and upper bounds for $ d_{m_1,m_2}  [k] $ in (3) are $ d_{m_1,m_2,min} [k]= $$ d_{m_1,m_2} [k-1]-(V_{m_1}^{max}+V_{m_2}^{max})\delta $ and $ d_{m_1,m_2,max} [k]= $$ d_{m_1,m_2} [k-1]+(V_{m_1}^{max}+V_{m_2}^{max})\delta $, respectively. Thus, the right-hand side in (5c) is lower bounded by
	
	\small
	\begin{gather}
		\sum_{k\in I_{m,M}} B_k  \text{log}_2(1+
		\frac{Q_{m,M,k}d_{M,m}^{-\alpha_{m,M}}}{\sigma_{m,M,k}^2+\sum\limits_{\substack{m'\in [1,M+1]\backslash{\{M\}}}} Q_{m,m',k}d_{m,m',min}^{-\alpha_{m,m'}}})\label{RHS-6c}
	\end{gather}%\vspace{-1\baselineskip}
	\normalsize	
	
	%	\begin{gather}
	%	(\ref{LHS-6c})\geq (\ref{RHS-6c})
	%	\end{gather}%\vspace{-1\baselineskip}
	\noindent Hence, we replace (5c) with the constraint (\ref{LHS-6c})$ \leq $(\ref{RHS-6c}).	Based on its derivation, once (\ref{LHS-6c})$ \leq $(\ref{RHS-6c}) is fulfilled, the constraint (5c) is fulfilled as well. 
	\textcolor{black}{Note that, a derivation of closed-form expression for $ d_{M,m} $ from (\ref{LHS-6c})$ \leq $(\ref{RHS-6c}) is difficult. Nevertheless, the term in (10) is a constant, and the term in (11) is a strictly decreasing function of $ d_{M,m}[k] $. Hence, we use bisection method to find an upper bound equivalent to (\ref{LHS-6c})$ \leq $(\ref{RHS-6c}) as} $ d_{M,m}[k]\leq D_{M,m}[k] $ for $ m\in [1,M-1] $. This inequality defines the border and interior of a sphere with a center at  $ \bm{q_{m}} $ and a radius of $ D_{M,m}[k] $. We refer to the combination of (5a), (5b), and (\ref{LHS-6c})$ \leq $(\ref{RHS-6c}) as the feasibility region, which results from intersections spheres (corresponding to (5b) and to (\ref{LHS-6c})$ \leq $(\ref{RHS-6c})) and the region between two planes (as in (5a)). 

Having investigated the impact of the constraints (5a)-(5c), we now focus on the approximated objective in (8).  According the approximated objective, for $ \rho_m>0 $, the $ m $-th FlyBS should move to the closest possible point to $ \bm{q_{0,m}} $ to maximize sum capacity. Similarly, for $ \rho_m<0 $, the $ m $-th FlyBS should move as far from $ \bm{q_{0,m}} $ as possible to maximize sum capacity. Therefore, the positioning of the FlyBSs at every time step is based on a minimization of the $ m $-th FlyBS’s distance from $ \bm{q_{0,m}} $ if $ \rho_m>0 $, or a maximization of the distance from $ \bm{q_{0,m}} $ if $ \rho_m<0 $ under the constraints (5a), (5b), and (\ref{LHS-6c})$ \leq $(\ref{RHS-6c}). Since the feasibility region is continuous if $ \bm{q_{0,m}} $ lays inside of the region, the closest possible point to $ \bm{q_{0,m}} $ is the point $ \bm{q_{0,m}} $ itself and the furthest point to $ \bm{q_{0,m}} $ is on the region’s border. If $ \bm{q_{0,m}} $ lays outside of the region, the closest/furthest point of this region to/from  $ \bm{q_{0,m}} $ is on the border of the region. This fact enables to find the optimal point by searching on the border of the feasibility region (if $ \bm{q_{0,m}} $ is already not the optimal). Since the feasibility region's border is either sphere, plane, or the intersection of those (which are circles), we thus search for the closest point to $ \bm{q_{0,m}} $ in the described candidate set which consists of spheres and planes and their intersection.
	
%\vspace{0\baselineskip}
	
		\subsection{Positioning of the relay }\label{ssec:3D}
	
	Once the FlyBSs' positions at the access link are updated according to the previous subsection III.B, the relay's postioning comes into effect. The relay’s movement is done considering (8) and (5d). Note that any movement of the relay would not violate  (5c), as all possible movements of the relay are already considered for the derivation of (\ref{LHS-6c})$ \leq $(\ref{RHS-6c}) in the previous subsection. 
	According to (8) the relay moves towards $ \bm{q_{0,M}} $ if $ \rho_M>0 $, or in the direction away from $ \bm{q_{0,M}} $ if $  \rho_M>0 $. In addition to the constraints (5a) and (5b), we also consider (5d) to guarantee the requirement on the backhaul capacity. 
	The terms in $ C_{M,m} $ and $ C_{M+1,M} $ (according to (3) and (4)) include the interference from other BSs that makes the constraint (5d) non-convex with respect to the relay’s position. To tackle this issue, in the following, we derive a convex constraint from (5d). By taking a similar approach as in the derivation of (10) and (11), the left-hand side in (5d) is upper bounded by 
		
\small
	\begin{gather}
		\sum_{m=1}^{M-1}\sum_{k\in I_{m,M}} B_k  \text{log}_2(1+\nonumber\\
		\frac{Q_{m,M,k}d_{m,M,min}^{-\alpha_{m,M}}}{\sigma_{m,M,k}^2+\sum\limits_{\substack{m'\in [1,M+1]\backslash{\{M\}}}} Q_{m,m',k}d_{m,m',max}^{-\alpha_{m,m'}}})\label{LHS-6d}
	\end{gather}%\vspace{-1\baselineskip}
	\normalsize

	\noindent By a similar approach, the right-hand side of (5d) is lower bounded by
	
	\small
\begin{gather}
	\small{\sum_{k\in [1,K]\backslash I_{M+1}} B_k  \text{\text{log}}_2(1+
	\frac{Q_{M,M+1,k}d_{M,M+1}^{-\alpha_{M,M+1}}}{\sigma_{M,M+1,k}^2+\sum_{m=1}^{M-1} Q_{M,m,k}d_{M,m,min}^{-\alpha_{m,M}}})\label{RHS-6d}}
\end{gather}%\vspace{-1\baselineskip}
\normalsize
	
	\noindent Thus, instead of (5d), we consider the constraint (\ref{LHS-6d})$ \leq $(\ref{RHS-6d}). 
The term in (12) is a constant, and the term in (13) is strictly decreasing with respect to $ d_{M,M+1} $. Hence, there exists  $ D_{M,M+1}$ such that (\ref{LHS-6d})$ \leq $(\ref{RHS-6d}) is equivalent to	$ d_{M,M+1}\leq D_{M,M+1} $, which demarcates the points on and inside a sphere centered at $ \bm{q_{0}} $ with a radius of $D_{M,M+1}$. The value of $ D_{M,M+1}$ is derived using bisection. According to (8) for $ \rho_M>0 $ ($ \rho_M<0 $) the relay moves to the closest (furthest) point to (from) $ \bm{q_{0,M}}$  fulfilling (5a), (5b), and (\ref{LHS-6d})$ \leq $(\ref{RHS-6d}). The optimal position is found similarly as in subsection III.B.
	
	\subsection{Association of users to the BSs }\label{ssec:3E}

	After the positioning of the FlyBSs, we update the associated set of users to each BS including the GBS and the relay. It is noted that, the capacity of each user is independent of the association of other users to BSs, as the signal and interference for each user (as in (2)) would be uniquely determined from that user's association. Hence, the problem of user association is solved using linear programming (LP).
	After updating the association of the users, the next iteration of the FlyBSs’ positioning is applied. This iterative method continues until there is no further change in the user association, or until a maximum number of iterations is reached. The proposed iterative solution is applied at every time step.

	\section{Simulations and results}\label{sec:5}

	In this section, we present models and simulations adopted for evaluation of our proposed solution, and we demonstrate the advantages of the proposal over state of the art schemes.
	
	\subsection{Simulation scenario and models}\label{ssec:sim_scenario}

	We assume two circular areas, one with a radius of 400 m and with the GBS at its center, and another area with a radius of 400 m and with its center 1600 m away from the GBS. 25\% of  all users are distributed in the first circular area and 75\% distributed in the second area. Within each area, half of the users move based on a random-walk mobility model with a speed of 1 m/s. The other half of the users are randomly distributed into six clusters of crowds. The centers of three of the clusters move at a speed of 1 m/s. Each user in those clusters moves with a uniformly distributed speed of [0.6, 1.4] m/s with respect to the cluster's center. Furthermore, the centers of the other three clusters move at a speed of 1.6 m/s with the speed of users uniformly distributed over [1.2, 2] m/s with respect to the center of their corresponding cluster. 
	 
	 A total bandwidth of 100 MHz with $ C= $120 channels of equal bandwidths are assumed. For the GBS and the relay, 20\% of the channels are allocated to directly serve the users (i.e., $ |I_M |=|I_{M+1} |=0.2C $), and the rest are allocated for backhaul connection. The backhaul bandwidth is split equally among the FlyBSs at the access link. A maximum transmission power of 37 dBm and 30 dBm is considered for the GBS and the FlyBSs, respectively \cite{Nikooroo2022TWC}. The noise power is set to -90 dBm. Pathloss exponents of $  \alpha_{n,m}= $2.8 and $  \alpha_{m_1,m_2}= $2.1 are adopted for BS-to-user and BS-to-BS channels, respectively.  An allowed altitude range of $ [H_{min,m},H_{max,m} ]= $[100, 300] m and a maximum speed of 25 m/s are assumed for the FlyBSs. The results are shown for $ M= $\{2,3,4,5\} FlyBSs. 
	Each simulation has a duration of 1200 seconds. The results are averaged out over 100 simulation drops.  
	
	We compare our proposal against the performance of the following schemes: \textit{i}) a two-hop  version of our proposed solution to FlyBSs’ positioning and user association. This model is simply derived from our original proposed model by treating the relay as a ground base station. In such case the positioning of the FlyBSs is done according to subsection III.B.,  \textit{ii}) state-of-the-art work in \cite{Qiu2020TCOM} where a \underline{m}aximization of \underline{t}otal \underline{n}etwork’s \underline{u}tility (referred to as MTNU) is provided in a two-hop relaying FlyBS network via FlyBSs’ positioning, user association, and resource allocation. The utility is defined as the sum of logarithm of user’s capacities. \textit{iii}) \underline{m}aximization of \underline{m}inimum \underline{u}ser’s \underline{c}apacity (referred to as MmUC), via FlyBS’s positioning and resource allocation published in \cite{Li2018}.

	\subsection{Simulation results}\label{ssec:sim_result}

In this subsection, we present and discuss simulation results. 
Fig. 2 illustrates the sum capacity achieved by different schemes for different number of users and for $ M= ${2,5}. According to Fig. 2, the sum capacity increases with the number of users in general for all schemes as there are more (orthogonal) channels used. However, if all available channels are used, the sum capacity would saturate as in such case, not all the users will be served by the FlyBSs and, hence, there is no further increase in the sum capacity of the network. This can be seen in the lines for $ M= $2  in Fig. 2. According to Fig. 2, the proposed solution enhances the sum capacity by 80\%, 15\%, 14\%, 15\%, and 17\% compared to the two-hop proposal for $ N= $ 200, 300, 400, 600, and 800, respectively, and by 109\%, 41\%, 35\%, 38\%, and 40\% compared to MTNU for $ N= $ 200, 300, 400, 600, and 800, respectively, and by 15\%, 17\%, 19\%, 20\%, and 23\% compared to MmUC for $ N= $ 200, 300, 400, 600, and 800, respectively.

\begin{table}
	\vspace{0.2cm}
	\begin{minipage}{0.47\linewidth}		
		
		\centering
		\includegraphics[width=1.8in]{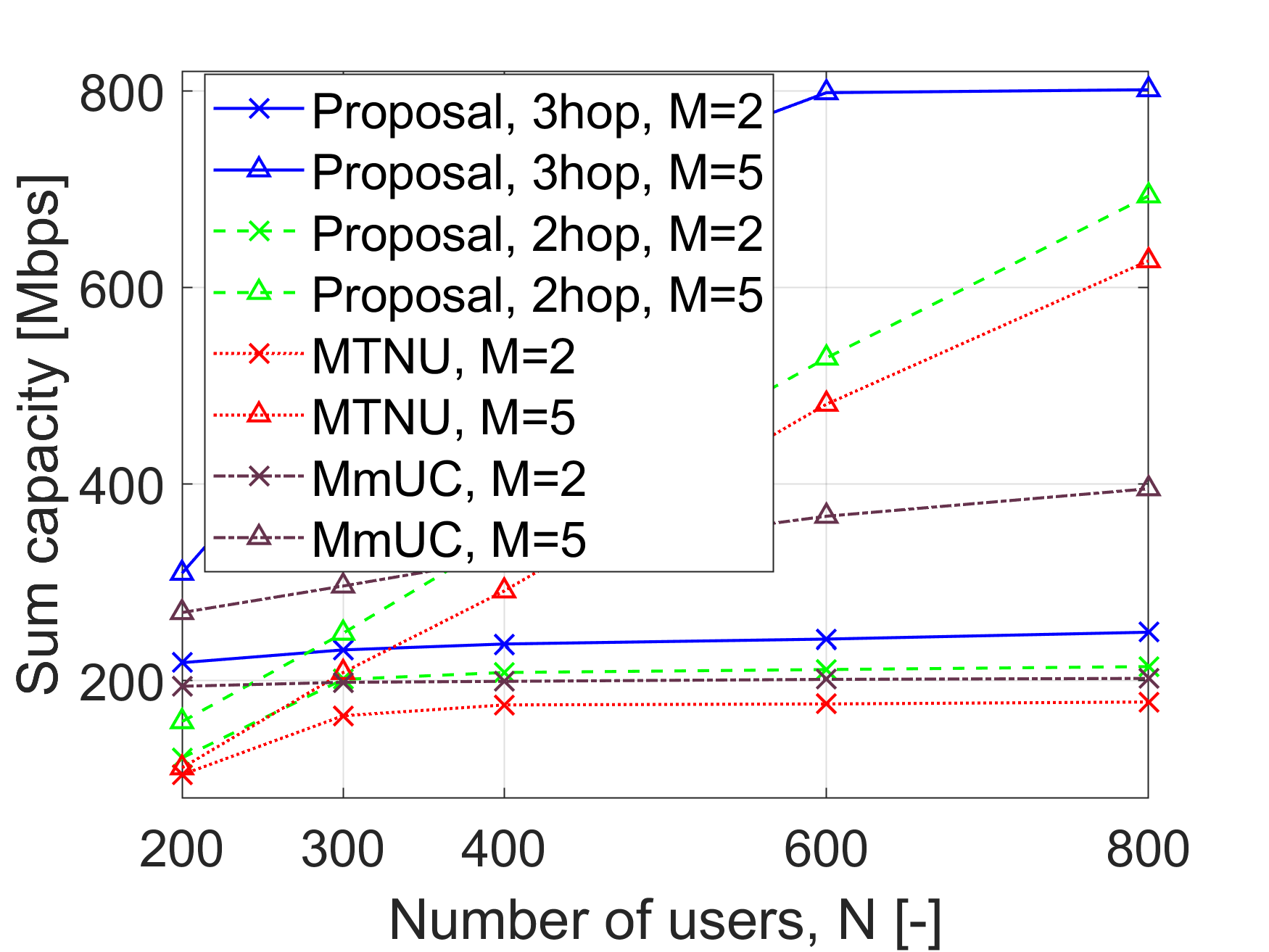}
		\captionof{figure}{\textcolor{black}{Sum capacity vs. $N$ for different schemes.}}
		\label{fig:9}
	\end{minipage}
	\hspace{0.01cm}
	\begin{minipage}{0.49\linewidth}\label{fig:CTO}	
		
		\centering
		\includegraphics[width=1.8in]{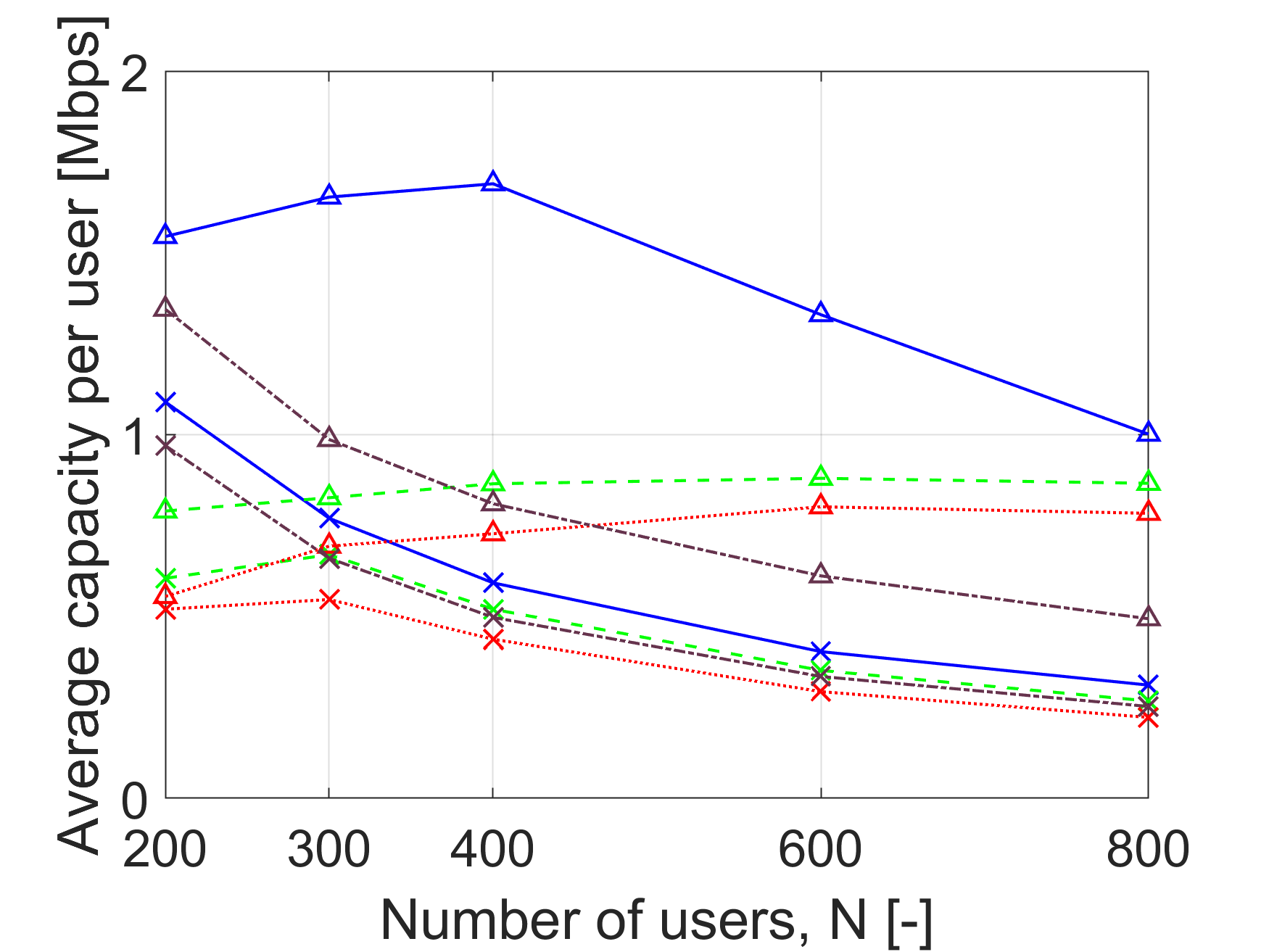}
		\captionof{figure}{Average user capacity vs. $N$ for different schemes. }
		\label{fig:CTO}
	\end{minipage}
\end{table}

Following the comparison in terms of sum capacity, we show the average user capacity in Fig. 3 for different number of users and different schemes. According to Fig. 3, the average user capacity increases with $ N $ in case that there are unused channels to be allocated to the added users. The two-hop proposal and MTNU with $ M= $2 show a strictly increasing average user capacity for $ N $ larger than 200.  However, the average user capacity might decrease for larger number of users in case that there are not enough unused channels to accommodate the added users. An example for such trend is presented in Fig.3 for MTNU with $ M= $2  and for the three-hop proposal with $ M= $5  for $ N $ larger than 300 and 400, respectively. Note that, the three-hop model has one FlyBS less than the two-hop model at the access link for the same value of $ M $, and so the three-hop model would run out of unused channels at lower values of $ N $ compared to two-hop models. Nevertheless, the proposed three-hop solution still outperforms other schemes (with the same $ M $).  According to Fig. 3, the proposed three-hop solution increases the average user capacity compared to the two-hop proposal by 80\%, 15\%, 14\%, 15\%, and 17\% for 200, 300, 400, 600, and 800 users, respectively. Furthermore, compared to MTNU, the proposed three-hop scheme increases the average user capacity by 110\%, 41\%, 35\% 38\%, and 40\%, respectively. Compared to MmUC, the proposed solution enhances the average user capacity by 13\%, 17\%, 20\% 21\%, and 23\%, respectively.
%	\begin{figure}[!t]
%	\centering
%	\includegraphics[width=3.4in]{figs/COMBINED_N_may22.png}	
%	\captionsetup{justification=centering}
%	\caption{Average user capacity vs. number of FlyBSs for different schemes.} 
%	\label{fig:sysmodel}
%\end{figure}\vspace{0\baselineskip}

\begin{figure}[!t]
	\centering
	\includegraphics[width=2.4in]{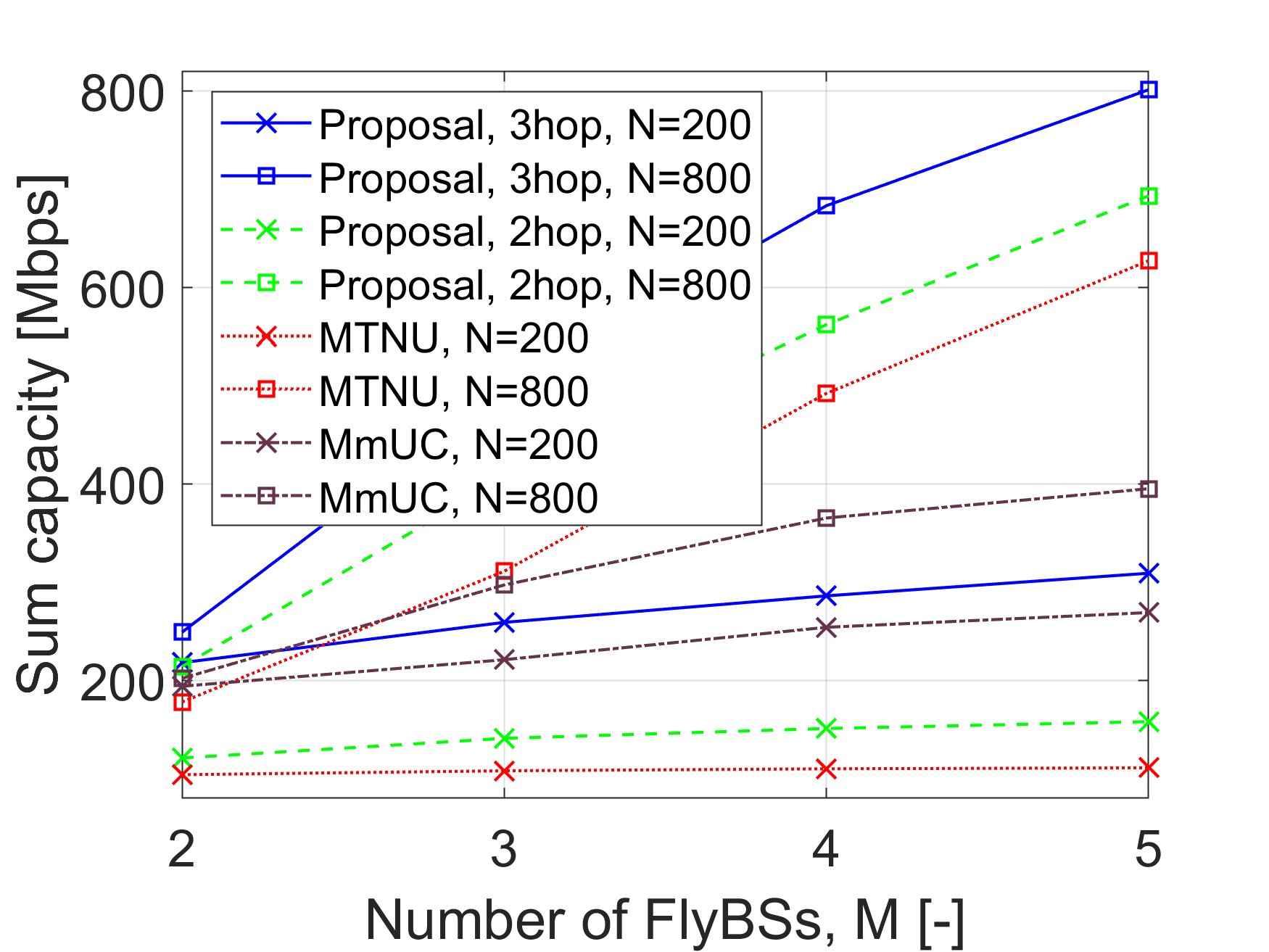}	
	\captionsetup{justification=centering}
	\caption{Sum capacity vs. number of FlyBSs for different schemes.} 
	\label{fig:sysmodel}
\end{figure}\vspace{0\baselineskip}

Next, Fig. 4 demonstrates the impact of the number of FlyBSs on the sum capacity for different schemes for $ N= $ 200 and $ N= $ 800. It is observed that, increasing the number of FlyBSs would increase the achieved sum capacity. This is mainly because of two reasons: 1) due to a limited number of channels, more number of users might be served by adding more FlyBSs, 2) even if all the users are already being served by the FlyBSs, adding more FlyBS might lead to a reassociation of some users to a closer FlyBS, resulting in a higher received power at the user despite an incurred interference by the other FlyBSs.
According to Fig. 4., for $ N= $200 the proposed solution increases the sum capacity by 80\%, 83\%, 90\%, and 95\% compared to two-hop proposal for M= 2,3,4, and 5, respectively, and by 109\%, 139\%, 160\%, and 178\% compared to MTNU for $ M= $ 2,3,4, and 5, respectively, and by 12\%, 17\%, 13\%, and 15\% compared to MmUC for $ M= $ 2,3,4, and 5, respectively. Furthermore, for $ N= $800, the proposed solution enhances the sum capacity by 17\%, 23\%, 21\%, and 16\% compared to two-hop proposal for $ M= $ 2, 3, 4, and 5, respectively, and by 40\%, 61\%, 39\%, and 28\% compared to MTNU for $ M= $ 2, 3, 4, and 5, respectively, and by 24\%, 68\%, 87\%, and 103\% compared to MmUC for $ M= $ 2,3,4, and 5, respectively.

	\section{Conclusions}
	
In this paper, we focus on multi-hop relaying FlyBS networks where there are FlyBSs adopted at both relaying and access links. We maximize the sum capacity with a consideration of backhaul constraints. To this end, we propose an analytical approach based on an alternating optimization of the FlyBSs’ positions and an association of users. The proposed solution improves the achieved sum capacity by tens of percent compared to existing solutions. In the future, the problem sum capacity maximization for multiple relays shall be studied.
	
	%\section{Acknowledgment}
	
	%\textcolor{red}{This work was supported by the project No. LTT 20004 funded by Ministry of Education, Youth and Sports, Czech Republic and by the grant of Czech Technical University in Prague No. SGS20/169/OHK3/3T/13.}

	%\footnotesize
	
	%\normalsize


\begin{thebibliography}{00}
		
		\bibitem{Liu2021} T. Liu, et al, "3D Trajectory and Transmit Power Optimization for UAV-Enabled Multi-Link Relaying Systems," {\em IEEE Trans. Green Commun. Netw.}, vol. 5, no. 1, 2021.
		
		\bibitem{Nikooroo2022TNSE} M. Nikooroo and Z. Becvar, "Optimization of Total Power Consumed by Flying Base Station Serving Mobile Users," {\em IEEE Trans. Netw. Sci. Eng.}, Early Access, 2022.
			\bibitem{Li2019} G. Li, et al, "Joint User Association and Power Allocation for Hybrid Half-Duplex/Full-Duplex Relaying in Cellular Networks," {\em IEEE Syst. J.}, vol. 13, no. 2, 2019.
		\bibitem{Li2020} B. Li, et al, "Joint Transmit Power and Trajectory Optimization for Two-Way Multi-Hop UAV Relaying Networks,"  {\em IEEE ICC Workshops}, 2020.
			\bibitem{Huang2020} Y. Huang, et al, "Bandwidth, Power and Trajectory Optimization for UAV Base Station Networks With Backhaul and User QoS Constraints," {\em IEEE Access}, vol. 8, 2020.
				\bibitem{Qiu2020Access} C. Qiu, et al, "Backhaul-Aware Trajectory Optimization of Fixed-Wing UAV-Mounted Base Station for Continuous Available Wireless Service," {\em IEEE Access}, vol. 8, 2020.
					\bibitem{Cicek2020} C. T. Cicek, et al, "Backhaul-Aware Optimization of UAV Base Station Location and Bandwidth Allocation for Profit Maximization," {\em IEEE Access}, vol. 8, 2020.
						\bibitem{Pham2021} Q. Pham, et al, "Joint Placement, Power Control, and Spectrum Allocation for UAV Wireless Backhaul Networks," {\em IEEE Netw. Lett.}, vol. 3, no. 2, 2021.
							\bibitem{Yu2021} Y.Yu, et al, "UAV-Aided Low Latency Multi-Access Edge  Computing,"  {\em IEEE Trans. Veh. Technol.}, vol. 70, no. 5, 2021.
						
		
		\bibitem{Qiu2020TCOM} C. Qiu, et al, "Multiple UAV-Mounted Base Station Placement and User Association With Joint Fronthaul and Backhaul Optimization," {\em IEEE Trans. Commun.}, vol. 68, no. 9, 2020.
		
	     	\bibitem{Mach2022} P. Mach, et al, "Power Allocation, Channel Reuse, and Positioning of Flying Base Stations With Realistic Backhaul," {\em IEEE Internet Things J.}, vol. 9, no. 3, 2022.
	     	
	     		\bibitem{Iradukunda2021} N. Iradukunda, et al, "UAV-Enabled Wireless Backhaul Networks Using Non-Orthogonal Multiple Access," {\em IEEE Access}, vol. 9, 2021.
	     			\bibitem{Pan2019} C. Pan, et al, "Joint 3D UAV Placement and Resource Allocation in Software-Defined Cellular Networks With Wireless Backhaul," {\em IEEE Access}, vol. 7, 2019.
	     	
		
	
		
		\bibitem{Li2018} P. Li and J. Xu, “Placement Optimization for UAV-Enabled Wireless Networks with Multi-Hop Backhaul”, {\em J.Commn.Net}, vol. 3, no. 4, 2018.
		
	
		
	
	
		
	
		%\bibitem{Youssef2020} M. Youssef, et al, "Full-Duplex and Backhaul-Constrained UAV-Enabled Networks Using NOMA," {\em IEEE Trans. Veh. Technol.}, vol. 69, no. 9, 2020.
		
	
		
	
	
		
	
		
		\bibitem{Sabzehali2021} J.Sabzehali, et al, “Optimizing Number, Placement, and Backhaul Connectivity of Multi-UAV Networks”, arXiv:2111.05457.
		
		
		
		\bibitem{Wang2020} L. Wang, et al, "An Integrated Affinity Propagation and Machine Learning Approach for Interference Management in Drone Base Stations," {\em IEEE Trans. Cogn. Commun. Netw.}, vol. 6, no. 1, 2020.
		
	
		
		\bibitem{Nikooroo2022TWC} M. Nikooroo and Z. Becvar, "Optimal Positioning of Flying Base Stations and Transmission Power Allocation in NOMA Networks," {\em IEEE Trans. Wireless Commun.}, vol. 21, no. 2, 2022.
		
		\bibitem{Esrafilian} O. Esrafilian, R. Gangula and D. Gesbert, "Learning to Communicate in UAV-Aided Wireless Networks: Map-Based Approaches," {\em IEEE Internet Things J.}, vol. 6, no. 2, 2019.
		
		
	
		
		
	\end{thebibliography}
\end{document}